\documentclass[twocolumn,prb,amsmath]{revtex4-2}

\usepackage{graphicx}
\usepackage{dcolumn}
\usepackage{bm}
\usepackage{color}
\usepackage{float}
\usepackage{hyperref}
\usepackage[normalem]{ulem}

\begin{document}

\title{Charge carrier relaxation dynamics in the one-dimensional Kondo lattice model}

\author{Arturo Perez-Romero}
\email{arturo.perezromero@uni-goettingen.de}
\affiliation{Institut f\"{u}r Theoretische Physik, Georg-August-Universit\"{a}t G\"{o}ttingen, D-37077 G\"{o}ttingen, Germany}
\author{Mica Schwarm}%
 \affiliation{Institut f\"{u}r Theoretische Physik, Georg-August-Universit\"{a}t G\"{o}ttingen, D-37077 G\"{o}ttingen, Germany}

\author{Fabian Heidrich-Meisner}
\email{heidrich-meisner@uni-goettingen.de}
\affiliation{Institut f\"{u}r Theoretische Physik, Georg-August-Universit\"{a}t G\"{o}ttingen, D-37077 G\"{o}ttingen, Germany}

\date{\today}

\begin{abstract}

A generic question in the field of ultrafast dynamics is concerned with the relaxation dynamics and the subsequent thermalization of optically excited
charge carriers. Among several possible relaxation channels
available in a solid-state system, we focus on the coupling
to magnetic excitations.
In this paper, we study the real-time dynamics of a paradigmatic model, the Kondo lattice model in one dimension. 
We conduct a comprehensive study of the relaxation processes by evaluating the spin polarization of the conduction electron, the local spin-spin correlation between localized and conduction electrons,  and the electronic momentum distribution. 
While in the well-studied cases of one or two charge carriers in a ferromagnetic background, no thermalization occurs, we demonstrate that the stationary state is compatible with thermalization 
if either the electronic filling is finite or the magnetic background is in the singlet sector. 
 Our real-time simulations using the time-dependent Lanczos method are corroborated by a direct comparison with finite-temperature expectation values and an analysis of the spectrum in terms of the gap ratio.

\end{abstract}

\maketitle

\section{\label{sec:level1}Introduction}

Theoretical research into the nonequilibrium dynamics of strongly-correlated quantum many-body systems 
continues to be fueled from several directions. First, the experimental advances with ultrafast dynamics
provide access to ever-more time windows and observables, unveiling physics that has been out of reach so far \cite{boschini2024, zhang2016, Giannetti2016, mitrano2016, dal2015, orenstein2012, fausti2011, yusupov2010,cavalleri2001}.
This includes the short-time dynamics in the electronic relaxation dynamics \cite{cavalleri2001,dal2015}, stabilization of metastable states \cite{zhang2016,yusupov2010},
and controlled driving of materials into states with a desired type of order \cite{mitrano2016,fausti2011}. 
Second, modeling the relevant materials often requires accounting for several degrees of freedom, such as
itinerant electrons, orbital degrees of freedom, phonons, or magnetic excitations, giving rise to a high degree of 
complexity \cite{basov2017,dagotto2005}.
Third, theoretical descriptions
face formidable challenges, since several decades in time need to be covered with different physics dominating different time windows \cite{wilner2014,aoki2014,Moshnyaga2025}. This is usually not possible without approximations, and therefore, simple but generic 
model Hamiltonians such as the Hubbard model  \cite{Murakami2025} or Kondo-lattice models \cite{Werner2012,zhu2021,chen2024} continue to play an important role.

In this work, we will take the route of working with a model system, and we are specifically interested in the relaxation dynamics of electrons 
coupled to a magnetic subsystem. This falls into a broader class of problems, where one seeks to unravel the electronic time evolution after an optical excitation due to a coupling to bosonic degrees of freedom \cite{Stolpp2020, Koehler2018, Dorfner2015, Golez2012}. The latter are usually either phonons or magnons. Several generic initial conditions have been studied: (i) the relaxation of a single charge carrier with a well-defined quasimomentum in an empty band \cite{Golez2012, Dorfner2015, Lenarcic2014}, 
(ii) the dynamics of a charge carrier in real-space \cite{Dorfner2015, Golez2012, Mierzejewski2011}, 
(iii) the relaxation  of one or many electrons after an optical excitation modeled as a pulse \cite{yusupov2010,Koehler2018, Murakami2025}, (iv) the specific analysis of the time evolution of an order parameter $\mathcal{O}(t)$ after such initial conditions, either concerning their decay as a function of time \cite{polkovnikov2011, Fauseweh2020, Stolpp2020,osterkorn22, Paprotzki2024} or the emergence of a nonzero $\mathcal{O}(t)$ due to the optical excitation \cite{Shirakawa2020, Paeckel2020, Fauseweh2025}.

\begin{figure}[t] 
\centering
\includegraphics[width=1.0\linewidth]{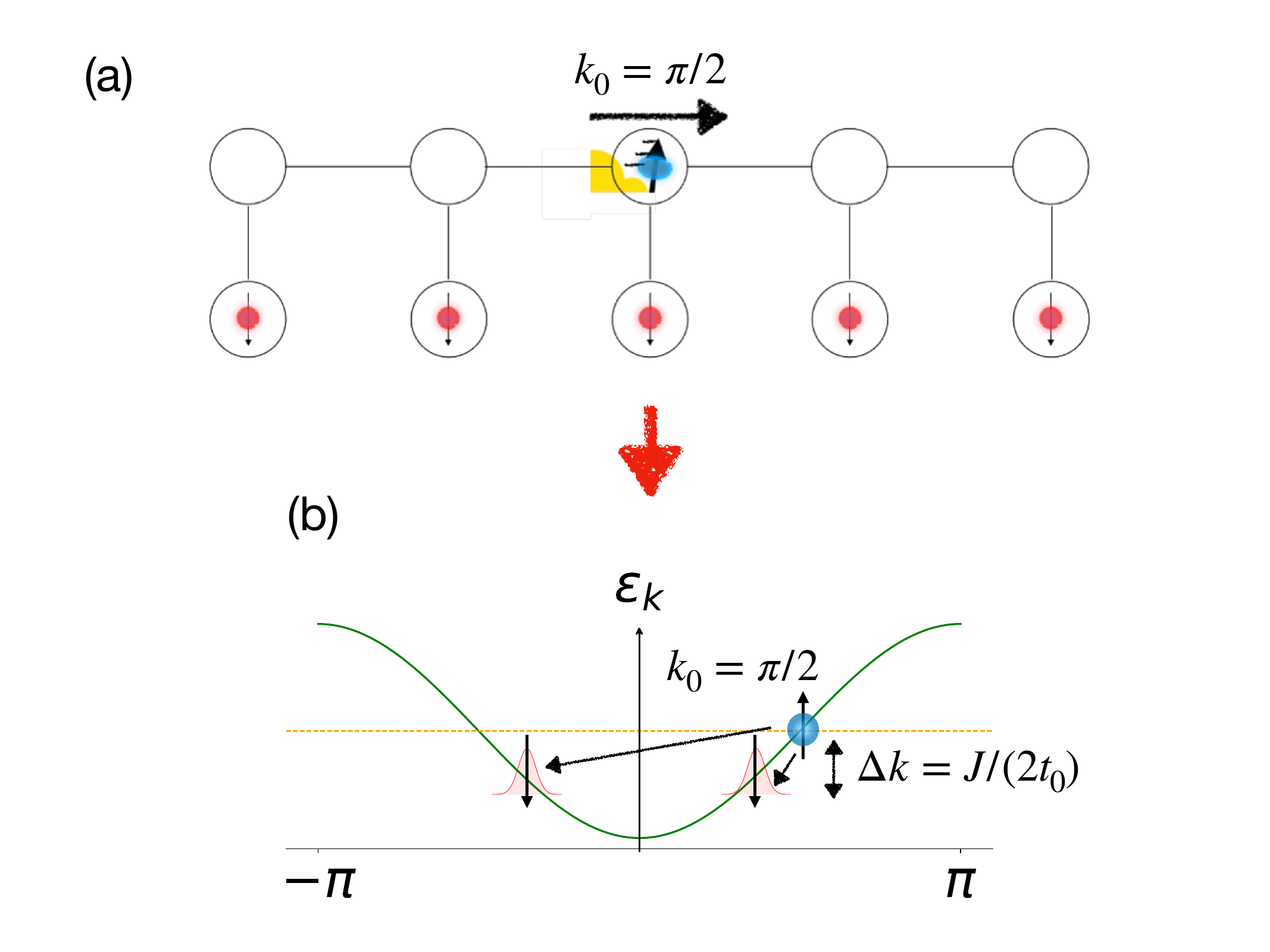}
\caption{(a)  Sketch of the initial condition, Eq.~\eqref{eq:initial_state1}, where the localized spins are initially fully polarized 
and the conduction electron is prepared in a quasimomentum eigenstate with quasimomentum $k_0$ and a spin orientation opposite to the one of the localized spins.  
(b) Illustration of scattering processes due to interactions with spin excitations of the localized electrons. $\Delta k=J/(2 t_0)$ is the momentum transferred to the spin excitations [see Eq.~\eqref{eq:Ham_spin}] for the initial state $|\psi_{0,F}\rangle$ in the limit of $J\ll t_0$.}
\label{fig:sketch}
\end{figure}

In our work, we start from the  first setup, which we will modify as outlined below. 
We pursue the strategy of obtaining numerically exact results, and therefore, we restrict the discussion to the one-dimensional case. As a model Hamiltonian,
we consider a generalized Kondo-lattice Hamiltonian \cite{ueda1991,tsunetsugu93,Tsunetsugu1997,gulacsi06,trebst2006,basylko08,smerat09,arredondo12} as sketched in Fig.~\ref{fig:sketch}(a), incorporating the Kondo coupling between localized spins and the itinerant electrons, a direct coupling between the spin degrees of freedom, and disorder. On the one hand, we aim at establishing a comprehensive understanding
of the relaxation dynamics depending on the model parameters and for  idealized initial conditions in the spin sector. On the other hand, we want to address the issue of thermalization.

 Both the cases of a coupling of an optically excited electron to phonons and magnons have been investigated. For the former, we mention \cite{Golez2012,sentef2016,park2022,caruso2022}, while for the latter, Refs.~\cite{Mierzejewski2011,Golez2014,mondal2019} are examples.
As an ultimate goal beyond the scope of the present study, we wish to understand the competition between several bosonic relaxation channels, e.g., in the presence of phonons and magnons (see Refs.~\cite{Kogoj2014,Kogoj2016} for work in that direction).

To set the stage, we paraphrase results from Ref.~\cite{Dorfner2015}, where a charge carrier was initialized in a quasimomentum eigenstate in an empty band and then coupled to optical phonons via a Holstein coupling. In the adiabatic regime where phonon frequencies $\omega_0$ are smaller than the electronic bandwidth, Ref.~\cite{Dorfner2015}  reports a fast relaxation to a quasistationary state. That quasistationary state is characterized by a complete energy transfer from the electron to the optical phonons accompanied by a relaxation of the average electronic momentum to zero. Interestingly, the final momentum distribution appears to be thermal already in a simple model such as the Holstein chain studied in Ref.~\cite{Dorfner2015} (substantiated by follow-up studies \cite{Jansen2019,Schoenle2021}). Moreover, the typical time scale for a complete momentum and energy transfer was extracted in dependence of relevant parameters such as the electron-phonon coupling and the oscillator frequency. In the present study, we aim at establishing similarities and differences between the relaxation dynamics in the Holstein model and the Kondo-lattice chain.

In this work, we consider initial states that are
product states between the two subsystems, the itinerant electron band (subscript $c$) and the localized electrons  (subscript $f$):
\begin{align}
|\Psi_{0}\rangle & = |\phi_0 \rangle_{c} \otimes | \psi_0 \rangle_f \label{eq:initial_state0}\,.
\end{align}
The initial state $|\phi_0\rangle_c$ for the electrons will consist of $N_e$ electrons prepared in well-defined quasimomentum states. For the spin part, we will consider product states such as a fully polarized state or states with an antiferromagnetic pattern.
Our focus will be on analyzing the average spin polarization of the itinerant electrons, their momentum distribution and kinetic energy, as well as local correlations between the itinerant electrons and the localized spins.

As a starting point, let us consider the well-studied case (see, e.g., \cite{Shastry81,sigrist91,Nakano2012,Henning2012,Moeller2013,Frakulla2024}) of 
a single charge carrier moving in the ferromagnetic background 
(short FM) of the
localized spins [see Fig.~\ref{fig:sketch}(a)], thus an initial state of the form:
\begin{align}
|\Psi_{0,F}\rangle & = |k_0=\pi/2, \uparrow\rangle_c \otimes | \text{FM} \rangle_f \label{eq:initial_state1}\,.
\end{align}
In this exactly solvable case \cite{sigrist91,Nakano2012,Henning2012} of effectively two particles---electron and one emitted magnon---the system relaxes, yet evidently not to
a thermal state, as evidenced by the persistence of finite-momentum peaks in the quasimomentum distribution.
Likewise, for the case of two charge carriers in a ferromagnetic
background, no thermalization is observed as these are still few-body problems in a subspace with subexponential
scaling with system size \cite{Moeller2012a,Moeller2012, Rausch2019}.
We will start by illustrating these behaviors using numerical simulations to provide a comprehensive basis for our study.
We further consider the addition of diagonal disorder seen by the conduction electron. As expected, Anderson localization leads to an additional suppression of thermalization.
However, at weak disorder, we still observe a complete magnetization transfer. For large disorder and a large $J$, the dynamics is purely oscillatory, dominated by local dimers.

From these limiting cases, we move the system towards generic many-body behavior in two ways, touching either the electronic $|\phi_0\rangle_c$ or the spin portion $|\psi_0\rangle_f$ of the initial state: (i) either we increase the particle number in a ferromagnetic background, thus mimicking a finite density of charge carriers, or (ii) we keep one charge carrier but now injected into a system with antiferromagnetic correlations. As a main result, in both these cases, we observe a dynamics at long times that is consistent with 
thermalization. To support this conclusion we study the gap ratio, the decay of the spin polarization of the charge carriers, and in particular, the time-dependence of the quasimomentum distribution function. The latter forms a broad distribution centered around $k=0$.

The plan of the paper is the following. In Sec.~\ref{sec:level2}, we introduce the model, define the observables under study, and describe the numerical methods employed. We also describe the application of perturbative techniques in limiting cases.  
Section~\ref{sec:level3.1} discusses the relaxation from
an initial state with fully polarized local spins for  a single charge carrier, with 
Sec.~\ref{sec:disorder} considering the situation with additional diagonal disorder. 
For the case of one charge carrier in the ferromagnetic background, we discuss the expected absence of thermalization in Sec.~\ref{sec:thermalization}.
In Sec.~\ref{sec:many-particles}, we increase the number of electrons but keep the fully polarized spin background. Section~\ref{sec:AFM}
covers a different initial state in a large many-body subspace,
for a single charge carrier.
 Finally, we present our conclusions in Sec.~\ref{sec:level4}.
Appendix~\ref{app:weight} discusses the decomposition of initial states and ground states into singlet and triplet sector for a single charge carrier.
 Appendix \ref{app:markov} summarizes the effective description of the small-$J/t_0$ regime for a single-charge carrier decay in a fully polarized background.

\section{\label{sec:level2}Models and methods}

In this section, we introduce the Hamiltonian and the set of observables that we will compute for different initial states and model parameters.
 Next, we describe the computational method used for the numerical diagonalization of the Hamiltonian and for time propagation. 
We further describe how we compute thermal expectation values of the same
observables. Finally, we derive a standard expression for short time scales.

\subsection{\label{sec:level2.1}Kondo-lattice model}

 The model is defined by the Hamiltonian:
\begin{align} \label{eq:kondo} 
{\hat H} =&-t_0 \sum^{L}_{i=1,\sigma}({\hat c}^{\dagger}_{i+1,\sigma}
{\hat c}_{i,\sigma}+H.c)+J\sum^{L}_{i=1} 
{\hat{\textbf{S}}}_{c,i} \cdot {\hat{\textbf{S}}}_{f,i} \nonumber \\
&+ J_{\text{AFM}} \sum^{L}_{i=1} {\hat{\textbf{S}}}_{f,i} \cdot {\hat{\textbf{S}}}_{f,i+1} + \sum^{L}_{i=1}\varepsilon_i {\hat n}_i\,.
\end{align}

The first term describes the kinetic energy of the itinerant electron.
In the following, we use a sub/super index $c(f)$ to refer to the itinerant(localized) electron(spins).  $t_0$ denotes the hopping matrix element. Here, $\hat c^{\dagger}_{i+1,\sigma}$ ($\hat c_{i,\sigma}$) represents the creation (annihilation) operator of the itinerant electron with spin $\sigma=\uparrow,\downarrow$ on  site $i$. The second term is the Heisenberg interaction between the spin degree of freedom of the itinerant electrons and the localized electrons,
governed by the exchange coupling strength $J$.  
${\hat{\textbf{S}}}_{c, i}$ and ${\hat{\textbf{S}}}_{f, i}$   are spin-1/2 operators describing the spin degree of freedom of the itinerant and localized electrons on site $i$, respectively. 
We express the Hamiltonian in terms of the  $z$-component ($\hat S^{z}_{c/f,i}$), the raising ($\hat S^{+}_{c/f,i}$), and lowering ($\hat S^{-}_{c/f,i}$) operators associated with $\hat{{\textbf{S}}}_{c/f,i}$. The third term represents a nearest-neighbor Heisenberg interaction between the localized spin degrees of freedom, with an  antiferromagnetic  exchange coupling $J_{\text{AFM}}$. The final term denotes a diagonal disorder 
experienced by the conduction electrons, where $\varepsilon_i$ is a local random potential drawn from the interval $\varepsilon_i \in  [-W, W]$.  $\hat n_i$ is the electron number operator on the site $i$ defined as $\hat n_i=\sum_\sigma \hat c^{\dagger}_{i,\sigma}\hat c_{i,\sigma}$.


We will start with the case  of a single itinerant electron, \textit{i.e.}, the filling is $n=1/L$, where $L$ is  the number of sites in a one-dimensional lattice. 
Then we will move on to two electrons with $n=2/L$, and finally to a finite density $n=1/4$.
We use periodic boundary conditions, allowing the use of a quasimomentum representation in the absence of diagonal disorder. 

In the absence of disorder and at $J=0$, the electronic part is diagonal in the quasimomentum basis
\begin{align} \label{eq:dis_tbm}
\hat H_{c}= \sum_{k,\sigma}\epsilon_k \hat c^{\dagger}_{k,\sigma} \hat c_{k,\sigma}\, , 
\end{align}
where $k$ is the single-particle quasimomentum chosen to lie in the first Brillouin zone, and the dispersion is (the lattice spacing is set to unity)
\begin{align}
\epsilon_k=-2 t_0\cos(k) \,.
\end{align}
It is illustrative to rewrite the term in the Hamiltonian 
that couples the spin of the conduction electrons to the
spin of the localized electrons in the quasimomentum basis.
We use standard definitions for the discrete Fourier transformation
\begin{align}
   \hat  c^{\dagger}_{k,\sigma} &= \frac{1}{\sqrt{L}}\sum_{j=1}^L e^{ijk} \hat c^{\dagger}_{j,\sigma}, \label{eq:cre_mom} \\
    \hat S^{\alpha}_{k}&= \frac{1}{\sqrt{L}}\sum_{j=1}^L e^{-ijk} \hat S^{\alpha}_j\, ,\label{eq:spi_mom} 
\end{align}
where $\alpha=z,+,-$.
The Hamiltonian term proportional to $J$ then takes the form:
\begin{align} \label{eq:Ham_spin}
\begin{split}
\hat H_{J} =& \frac{J}{2\sqrt{L}}\sum_{k,q}S^{z}_{f,q}  \left\lbrack\left(\hat c^{\dagger}_{k,\uparrow} \hat c_{k+q,\uparrow} - \hat c^{\dagger}_{k,\downarrow}\hat c_{k+q,\downarrow}  \right)\right.\\
&\left. +\hat S^{+}_{f,q}c^{\dagger}_{k,\downarrow}\hat c_{k+q,\uparrow}+ \hat S^{-}_{f,q}\hat c^{\dagger}_{k,\uparrow} \hat c_{k+q,\downarrow}\right\rbrack\,,
\end{split}
\end{align}
where the matrix element between the 
quasimomentum components of the localized spins and of the conduction electron becomes evident and is equal to  $J/[2\sqrt{L}]$.

\subsection{\label{sec:level2.2} Observables}

We study the following operators. First, we define the spin polarization of the itinerant electron as

\begin{align} \label{eq:spin_z_c} 
S^{z}_{c}(t)= \left< \hat S^{z}_{c}(t)\right> = \frac{1}{2}\sum^{L}_{i=1}
\left< \Psi (t) \right|(\hat{n}_{i,\uparrow} -\hat{n}_{i,
\downarrow})   \left| \Psi (t) \right>\,.
\end{align}

In our case, the initial value  is always $S^{z}_c(0)=1/2$. 
The second key quantity is the quasimomentum distribution function of the 
itinerant electron with spin orientation $\sigma$, whose expectation value is given by:
\begin{align} \label{eq:momentum_occ} 
\begin{split}
 n_{k,\sigma}(t) =&  \left< \Psi (t) \right|\hat c^\dagger_{k,\sigma}\hat c_{k,\sigma}   \left| \Psi (t) \right>  \,.
 \end{split}
 \end{align}
 We further define the total quasimomentum distribution as
\begin{align}
n_k(t) = \sum_{\sigma=\uparrow,\downarrow}n_{k,\sigma}(t) \,.
\end{align}
 From the quasimomentum distribution, we obtain the kinetic energy of the conduction electron:
 \begin{align}
 E_{\text{kin}}(t) = \sum_k \varepsilon_k n_k(t)\,.
 \end{align}
The final observable of interest is the correlator between the electronic spin density on site $i$ and the localized spin 
on site $i$:
\begin{align}
    S_{cf,i} & = \langle  {\hat{\textbf{S}}}_{c,i} \cdot {\hat{\textbf{S}}}_{f,i}\rangle\,.
\end{align}
Often, we will calculate the total correlation by summing over all sites:
\begin{align}
S_{cf} = \sum_{i=1}^L S_{cf,i} \,.
\end{align}
This accounts for the decreasing filling as $L$ increases in the case of a single charge carrier.

\subsection{\label{sec:level2.3} Numerical Method}

We use the time-dependent Lanczos method \cite{Manmana2005} to simulate the time evolution
of our many-body system and full exact diagonalization to compute thermal expectation values \cite{fehske2007,sandvik2010}.
We use periodic boundary conditions in all simulations.

We exploit symmetries, including total magnetization $S^z_{\text{total}}$, conservation of the number of conduction electrons $N_c$, and translational invariance  \cite{sandvik2010}.
The latter symmetry arises from the invariance of the Hamiltonian under lattice translations, i.e., eigenstates of $\hat H$ can also be chosen as eigenstates of the lattice translation operator $\hat T$, with eigenvalues $e^{iK}$, where the crystal momentum $K=2\pi n/L$ ($n=0,1,\ldots,L-1$) labels the different momentum sectors.  

In general, the dimension of the Hilbert space for a Kondo-lattice model with $L$ sites, $N_c$ conduction electrons, and total magnetization $S^z_{\text{total}}$ can be written as a sum over all possible spin-up electron configurations as
\begin{align}\label{eq:DimH} 
\text{dim}(\mathcal{H}) = \sum^{N_c}_{N_\uparrow=0} \left[ \binom{L}{N_\uparrow} \binom{L}{N_c - N_\uparrow} \binom{L}{L_\uparrow} \right] \,.
\end{align}
Here $N_{\uparrow}$ denotes the number of spin-up conduction electrons, and
\begin{align}
L_{\uparrow} = \frac{1}{2}\left[
L+2S^z_{\text{total}} -2N_\uparrow+N_c
\right]
\end{align}
is the number of localized spins with $S^z_\ell=1/2$. When the translation invariance is imposed, the size of the Hilbert space is reduced by roughly a factor of $L$.
This reduction is exact when all the representative states have a periodicity equal to the full system size. This condition holds for the initial state $|\Psi_{0,F}\rangle$, two electrons with opposite spin ($|\Psi_{0}^{(\uparrow\uparrow)}\rangle$), and the chosen state for the Kondo-Heisenberg model ($|\Psi_{0,KHM}\rangle$), discussed in Sec.~\ref{sec:AFM}.
As a special case, the dimension of the relevant Hilbert space is dim$(\mathcal{H})=L+1$ for the initial state $|\Psi_{0,F}\rangle$ since in that case, total  $S^z_{\text{total}}=L/2-1$, and we exploit translational invariance. 

In the time-dependent Lanczos method, we approximate the time-evolution operator $e^{-i\hat Ht}$ 
in a Krylov subspace of typically a small dimension
(we set $\hbar=1$). 
When applied to an initial state $\left| \Psi(t+\Delta t)\right>$, we obtain the expression
\begin{align} \label{asadas} 
 | \Psi(t+\Delta t)\rangle  \approx V^{\dagger}_{D}e^{-i\Delta t T_{D}}V_{D}| \Psi(t) \rangle,
\end{align}
where  $V_{D}$ maps from the original basis onto the Krylov basis and vice versa, $\Delta t$ is the  time step, and $T_{D}$ is a tridiagonal representation of the Hamiltonian $\hat H$ in the Krylov subspace.

For the propagation by one time step, we start by finding the Krylov subspace by iteratively applying the Hamiltonian to the current state $\left| \Psi(t)\right>$ $D_K-1$ times, 
and construct  the operator $T_{D}$ in the resulting Krylov basis. We find that $D_{K}=20$ is usually sufficient, as we verified by varying $D_K$. To diagonalize the matrix, we use the LAPACK package \cite{anderson1999} and a self-written code for the matrix multiplication to get the time-evolved state
 $\left| \Psi(t+\Delta t)\right>=V^{\dagger}_{D}e^{-i\Delta tT_{D}}V_{D}\left| \Psi(t)\right> $. 
 Next, we use the quantum state $\left| \Psi(t+\Delta t)\right>$ to compute expectation values of  operators at $t+\Delta t$ time.  Additional parameters used in this method are the time step, which has a range between $\Delta t/t_0 = 0.005$ and $\Delta t/t_0 = 0.05$ (depending on the regime and initial conditions), and a maximum system size of $L =226$ for the initial state $|\Psi_{0,F}\rangle$.

\subsection{Thermal expectation values}

We will compare the time-averages of expectation values $\langle \hat O(t) \rangle$ in the steady state to thermal expectation values in the canonical ensemble.
The thermal expectation value of the energy is 
\begin{align}
E = \mbox{tr} [ \rho \hat H] \,,
\end{align}
where $\rho$ is the statistical operator 
\begin{align}
\rho = \frac{1}{Z} \mbox{tr} [e^{-\beta \hat H}] \,.
\end{align}

Here, $\beta = 1/T$ is the inverse temperature (we set $k_B=1$) and $Z$ is the partition function.

Assuming thermalization \cite{rigol08,DAlessio2016},  the expected temperature is obtained  by equating the energy in the initial state to the thermal expectation value:
\begin{align}
E_0= \langle \Psi_0 | \hat H | \Psi_0 \rangle =  \mbox{tr} [ \rho \hat H]\,.
\end{align}
The thermal expectation value of a given observable  compatible with the initial energy $E_0$ is then given by
\begin{align}
\langle \hat O \rangle_T = \mbox{tr} [ \rho \hat O] \,.
\end{align}

 \subsection{\label{sec:level2.4} Short-time evolution}

It is instructive to determine the short-time evolution analytically. To that end, we expand the time-evolution operator $\hat U(t)=\mbox{exp}(-i\hat Ht)$. However, our Hamiltonian is composed of  non-commuting terms. Using the Baker-Campbell-Hausdorff (BCH) formula \cite{varadarajan2013}, we get the relation
\begin{align} \label{eq:short_time} 
e^{\hat X}\hat Oe^{-\hat X} = \hat O+[\hat X,\hat O]+\frac{1}{2!}[\hat X,[\hat X,\hat O]]+\cdots,\end{align}
where $\hat X=i\hat Ht$, and $\hat O$ is an observable. The error in this expansion is of the order of $t^n$, where $n$ is the number of terms  kept in the BCH expansion. We go up to second-order, leading to a good approximation for  $t \ll 1/t_0$. 

For the observables of interest in our study, we quote the  leading terms up to quadratic order in time.
For the spin polarization of the conduction electron, this reads:
\begin{equation}  \label{eq:small-time_Sz} 
S^z_{c}(t)  \approx  \frac{1}{2} - \frac{J^2}{4}t^2 \,.
\end{equation}
For the local spin correlation $S_{cf}$, we find
\begin{equation}  \label{eq:small-time_Ss} 
S_{cf}(t)  \approx  S_{cf}(t=0) - \frac{Jt_0}{2}t^2\cos({k_0}),
\end{equation}
where $k_0$ denotes the initial electronic quasimomentum. For $k_0=\pi/2$, the leading term vanishes. For other choices  of $k_0$, we verified that our numerical data agrees with this expression  for short times $Jt<0.2$ (not shown).

\section{Relaxation dynamics of a single charge carrier  in a ferromagnetic background}

\label{sec:level3.1}

We first focus on the case of an initial state of the type $|\Psi_{0,F}\rangle$ as defined in Eq.~\eqref{eq:initial_state1}, i.e.. the itinerant electron is prepared in a single-particle quasimomentum 
state with $k_0=\pi/2$ and spin up, while the localized spins are in a fully polarized state with a global
$S^z_{\text{total}}=(L-1)/2$. We will consider the case of $J_{\text{AFM}}=0$. The discussion will first cover the case without disorder and then we also present the relaxation dynamics in the presence of disorder.

The case without disorder has been solved  exactly  and was investigated from many perspectives in the literature \cite{sigrist91,Nakano2012,Henning2012}. We provide a discussion of this
case to set up the study of the many-body case,
putting the focus on a qualitative picture and
the behavior of the quasimomentum distribution function.

In Ref.~\cite{Henning2012} (extending earlier work, see Ref.~\cite{sigrist91}) all 
eigenstates on finite systems were constructed. The following
results were obtained for the time-dependence
of $S_c^z$: (i) the short-time dynamics of the spin polarization is
quadratic in time (which is generic), (ii)
in an intermediate time-window, the decay is exponential, (iii) the long-time behavior deviates from an exponential decay due to the van-Hove singularity at the bottom of the band, and (iv) oscillations occur at large $J/t_0$, The oscillations can be interpreted as an indication
of spin-polaron formation (see, e.g., \cite{Moeller2013}).

Several aspects of the relaxation dynamics in the regime of $J\ll t_0$ 
can be understood by considering the relaxation of a two-level system coupled to a generic bosonic bath, for which there is a substantial amount of literature \cite{seke1983,golosov1999,berman2010,crowley2022}.
Essentially, the conduction electron's polarization is transferred
in a single spin-flip process by creating an excitation in the spin bath, thus preventing the electronic kinetic energy from further relaxation. 
After a short coherent dynamics, there is a Markovian regime with an exponential decay with a time scale $J^2/t_0$. For $J/t_0\lesssim 1.5$, we find an excellent agreement between numerical data and these approximate analytical results. 

For $J/t_0\gtrsim 2$, eventually, 
on average, no net
energy transfer  occurs and the dynamics are mostly dominated by oscillations.
The steady-state values of expectation values can be understood
from considering the decomposition of the initial state into eigenstates in small systems, due to the local nature of the physics in the large $J/t_0>2$ limit.

\subsection{Numerical illustration of relaxation dynamics}

The initial state with a complete parallel alignment of all localized spins and antiparallel spin orientation of the conduction electron belongs to the subspace $S^z_{\text{total}}=\frac{1}{2}(L-1)$.  In principle, this coincides with the total spin quantum number of the ground state of the Kondo-lattice model with one itinerant electron. However, in the ground state, the lowest energy is obtained in the momentum sector $K=0$ when the conduction electron has the momentum $k_0=0$, implying that our initial state is well-separated from the ground state \cite{sigrist91}.
For small $J/t_0$ (in the range $0.1<J/t_0<2.0$), the initial state becomes essentially orthogonal to the ground state, with its full spectral weight residing in the excited states, as illustrated in Appendix~\ref{app:weight}.

 Figure~\ref{fig:ferro} shows the relaxation dynamics of the different expectation values for the initial state $|\Psi_{0,F}\rangle$ calculated by using the iterative Lanczos algorithm for $L=216$. We use a Krylov space of $D_K= 20$ and a time step of $\Delta t /t_0= 0.005$ throughout this section.
 By comparing with exact diagonalization, we conclude that these values of the control parameters are sufficient to suppress the numerical errors below $10^{-9}$. In order to elucidate finite-size effects, we study the relaxation dynamics for different system sizes and for different values of $J/t_0$. We observe that system-size dependencies in the time evolution depend significantly on the value of $J/t_0$. For $L=216$, the finite-size effects typically appear around $t \times t_0 \sim 100$ for small $J/t_0$ and around $t \times t_0 \sim 80$ for large $J/t_0$ (results not shown here).

\begin{figure*}[t]
\includegraphics[width=0.9\linewidth, height=8.5cm]{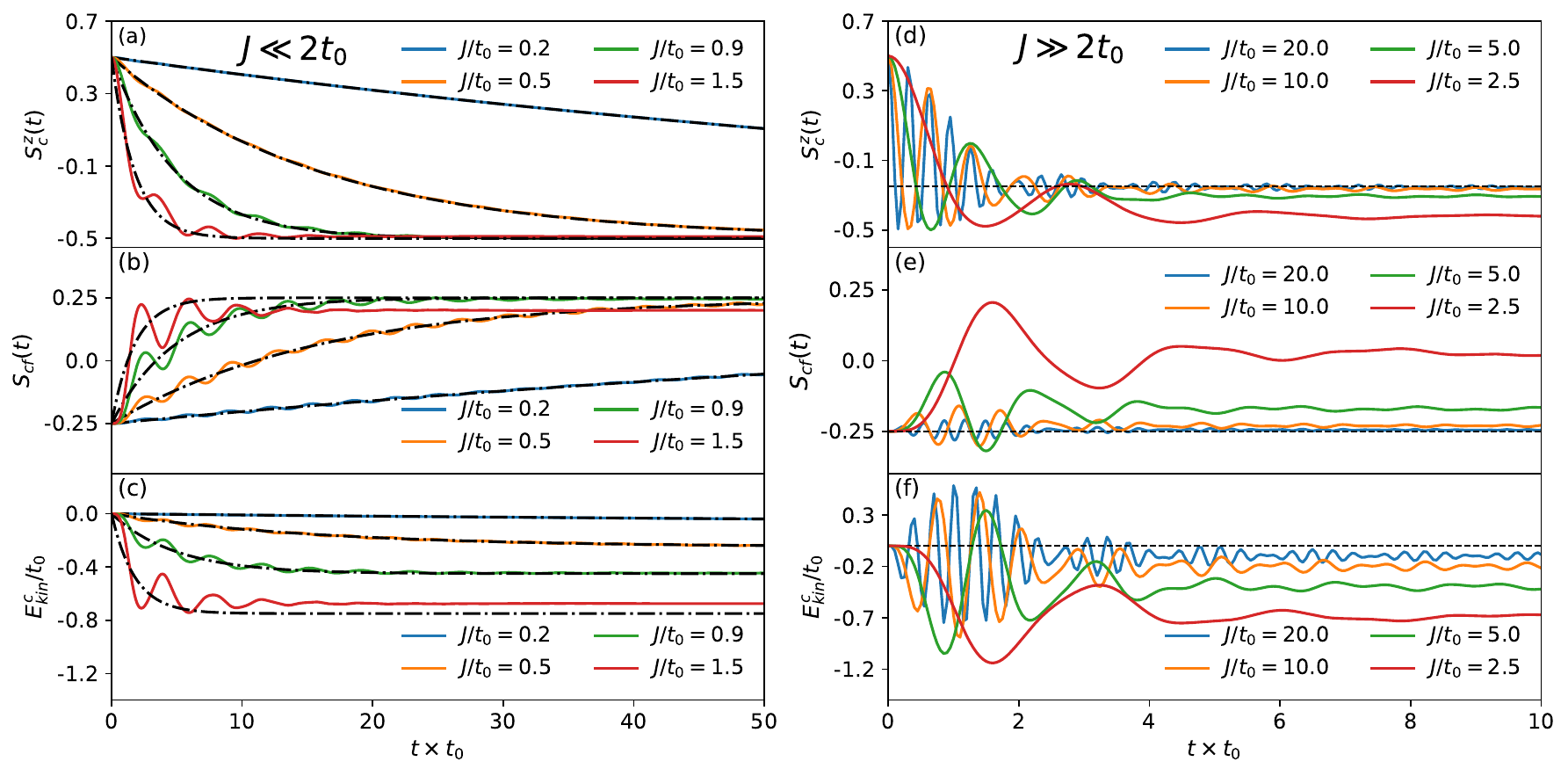}
\caption{\label{fig:ferro} Time evolution of several expectation values for various values of $J$ obtained numerically (solid lines) for the initial state $|\Psi_{0,F}\rangle$. (a)--(c): On the left side, we plot the results obtained for small $J/t_0=0.2, 0.5, 0.9, 1.5$, along with the results from the two-levels-plus-spin-bath (TLS) approximation (dash-dotted line, see Sec.~\ref{sec:TLS}). (d)--(f): On the right side, we present the results for large $J/t_0= 2.5, 5.0, 10.0, 20.0$. In (a) and (d), we show the  spin polarization of the conduction electron, $S_{c}^z(t)$. In (b) and (e), we display the total  local spin correlation $S_{cf}(t)$ between the conduction electron and the localized spins.  (c), (f): Kinetic energy of the conduction electron. The dashed lines in (d)--(f)  represent the 
(asymptotic)
long-time average for $J\gg 2t_0$ for the corresponding expectation value. These values are found to be: $S^z_c(t \rightarrow \infty) = -0.25$, $ S_{cf}(t \rightarrow \infty)= -0.25$, and  $E^c_{\text{kin}}(t \rightarrow \infty)/t_0=0.0$. } 
\end{figure*}

\subsubsection{ Regime of $J \ll 2t_0$}
\label{sec:level3.1.1}
As shown in Fig.~\ref{fig:ferro}, two separate regimes emerge based on the ratio $J/t_0$. 
We first discuss the regime when the electronic bandwidth 
exceeds the exchange coupling, i.e., $J<2t_0$. 
In this parameter regime, the expectation values of the operators exhibit a smooth time evolution, accompanied by small oscillations.
For the smallest Kondo coupling  shown in the Fig.~\ref{fig:ferro}, all expectation values undergo a slow relaxation dynamics. 
By increasing $J$, the spin polarization of the conduction electron decays more rapidly and the oscillation amplitude becomes larger, while the oscillation frequency remains practically unchanged. 

In addition, the $z$-component of the 
conduction electron $S_c^z$   and the total local spin correlations $S_{cf}$ approach a steady state without any apparent $J$-dependence. On the one hand,   $S_c^z$ [see Fig.~\ref{fig:ferro}(a)] approaches $S^z_c \to -0.5$, i.e., the electronic spin flips entirely from  $1/2$ to $-1/2$, becoming parallel to the initial spin orientation of the localized spins. On the other hand, 
$S_{cf} \to  0.25$ [see Fig.~\ref{fig:ferro}(b)]. Contrary to the expectation values of these two operators, the long-time behavior of the electronic kinetic energy, shown in Fig.~\ref{fig:ferro}(c), retains a clear dependence on the Kondo coupling $J$. To be specific, the total energy transfer to the local spin chain scales linearly as $\Delta E^{c}_{\text{kin}}=-J/2$.

\begin{figure}[t]
\includegraphics[width=0.95 \linewidth, height=9cm]{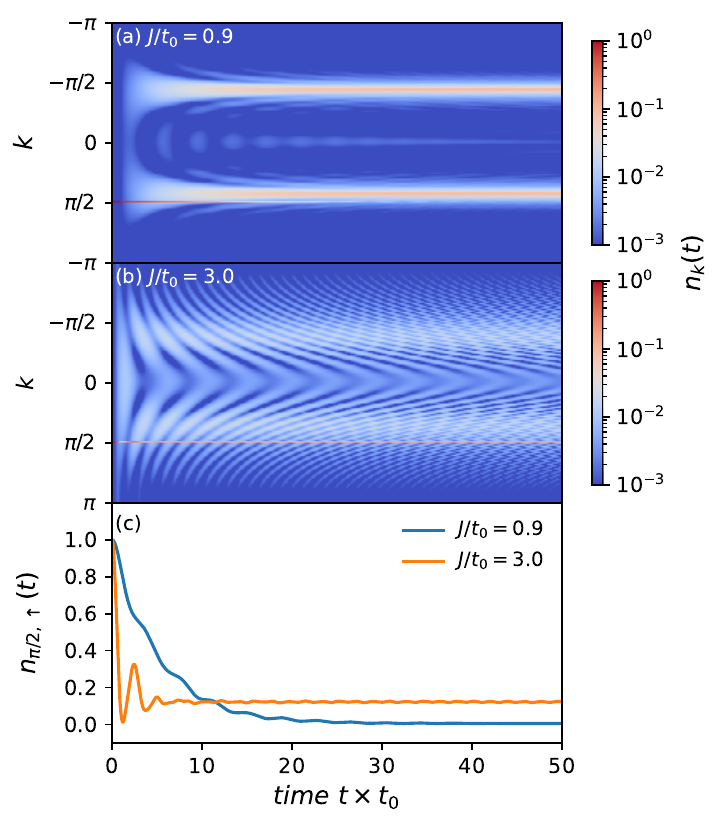}
\caption{\label{fig:mom_ferro} (a), (b): Logarithm of the electronic quasimomentum distribution $n_k(t)$ as a function of  time for a small ($J/t_0=0.9$) and a large ($J/t_0=3.0$) Kondo coupling, respectively. After a fast initial dynamics, $n_k(t)$ exhibits two well-defined peaks at $k=-\pi/2 + J/(4t_0)$, and $k=\pi/2 - J/(4t_0)$ for the case of  (a).
These finite-momentum peaks are inconsistent with a thermal state in the steady-state regime.
In (c), we illustrate the time evolution of $n_{k,\uparrow}(t)$ at momentum $k=\pi/2$ for both values of $J$. We set $L=216$, $k_0=\pi/2$, and an initial ferromagnetic configuration for the localized spins, i.e., $|\Psi_{0,F}\rangle$. }
\end{figure}

To complement the description of the small $J/t_0$ regime, we plot  the time-evolution of the electronic quasimomentum distribution function $n_k$ for  $J/t_0=0.9$ in Fig.~\ref{fig:mom_ferro}(a). The conduction electron starts with  $k_0 = \pi/2$ due to our initial condition from Eq.~\eqref{eq:initial_state1}. At short times,
$n_k$ retains a peak at  $\pi/2$ which then quickly decays. The momentum occupations redistribute into two distinct peaks: one just below $\pi/2$ and  another one above $-\pi/2$, located specifically at $k=\pi/2-J/(4t_0)$ and $k=-\pi/2+J/(4t_0)$. Since these peaks are sharp and far away from $k=0$,  the steady state does not correspond to a thermal state. This conclusion is further supported by a direct comparison between the thermal state expectation values and the stationary state values of the kinetic energy, discussed in
Sec.~\ref{sec:thermo}.

\begin{figure}[t]
\includegraphics[width=0.8\linewidth]{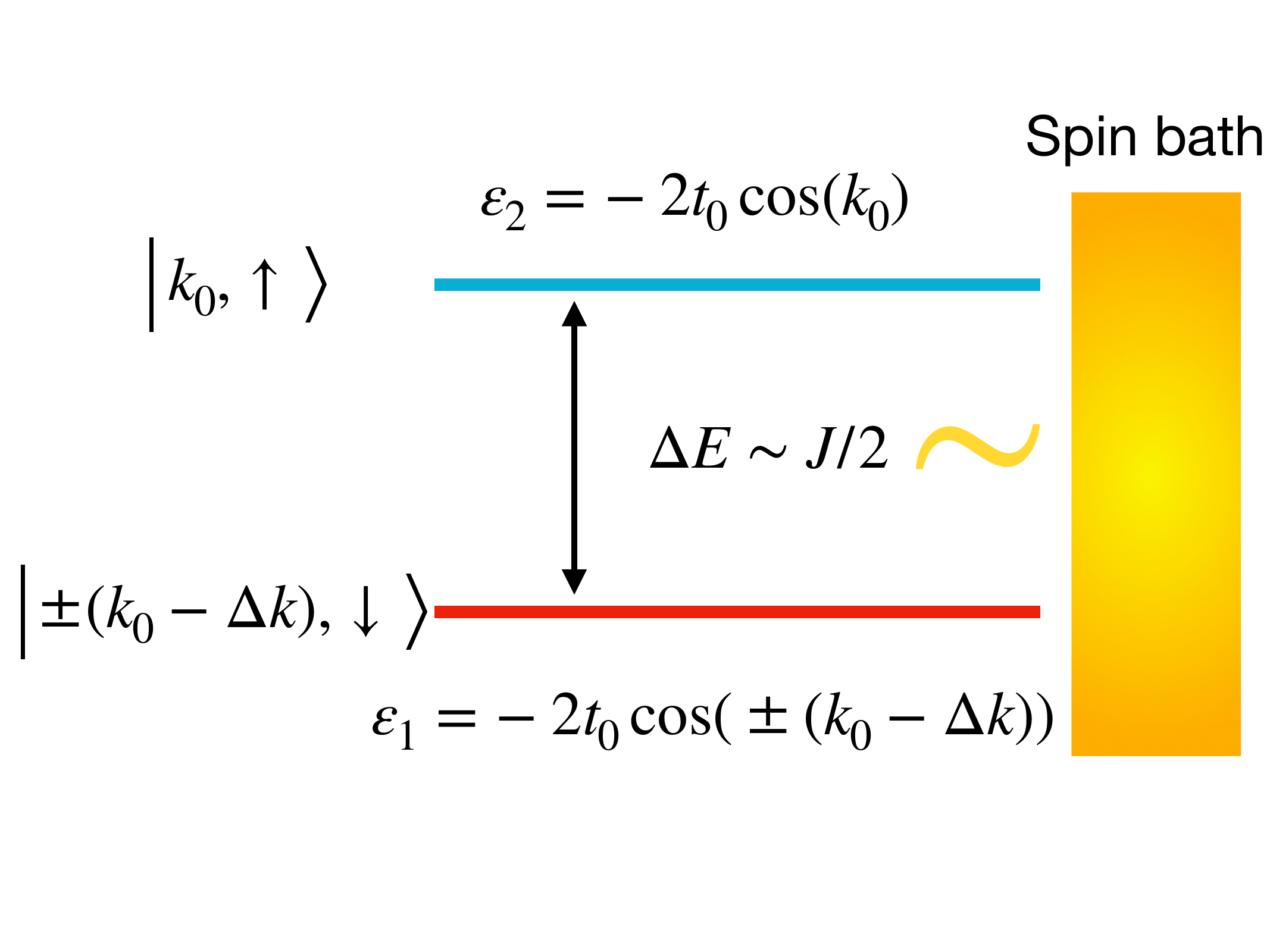}
\caption{\label{fig:TLS} Sketch of the effective two-level system    for the initial state Eq.~\eqref{eq:initial_state1} and for $J< t_0$. In this regime, there are only two relevant energies: the conduction electron with spin-up with energy $\varepsilon_2$ and with spin-down and energy $\varepsilon_1$. $\varepsilon_1$ is doubly degenerate.
The probability amplitude for the electron to be in the spin-up state is given by $b_2$, and for the spin-down state, it  is given by $b_{1,q}$ (see Appendix~\ref{app:markov}). 
Between both states, a spin-flip process occurs, emitting a magnon into the  spin bath with a total energy $\Delta E$. 
}
\end{figure}

Based on these observations, we conclude that, in the limit of $J \ll 2t_0$, the conduction electron 
undergoes a scattering process with four components: the electron acquires  a backward-propagating ($bp$) and  a forward-propagating ($fp$) component, and the same for the emitted magnon \cite{mitrofanov2020,tramsen2021}. After this scattering process, the $z$-component of the  electron spin is spin-down and  $k_{bp}=-k_{fp}$. Due to momentum conservation, the  magnon is excited  with weight at the opposite momenta $ q^m_{bp}=-q^{m}_{fp} $
(up to umklapp scattering).

\subsubsection{Two-level picture for $J/t_0\ll 1$}
\label{sec:TLS}

In Fig.~\ref{fig:TLS}, we show a sketch of the relevant processes for small $J/t_0$. There, we consider a simple two-level system (TLS) interacting with a bosonic bath. In this analogy, the bosonic bath represents the localized spins, while the spontaneously emitted excitation corresponds to the magnon propagating through the spin chain. The two relevant electronic levels  are the excited state $\left|e\right>$ and the ground state $\left|\textsl{g}\right>$.

For such a situation, standard approximations 
(see Appendix~\ref{app:markov}) yield
a Markovian behavior and hence an exponential
decay is found \cite{prokof2000}, in agreement with the exact solution \cite{Henning2012}. We extract the decay 
constant from a fit to the numerical data, resulting in  these functional dependencies for the  expectation values:
\begin{subequations}
    \begin{align}
    S^{z}_c(t)= -\left(\frac{1}{2}-e^{-\frac{J^2}{4t_0}t} \right), \label{eq:TL_8a}\\
    S_{cf}(t)  = \frac{1}{2}\left(\frac{1}{2}-e^{-\frac{J^2}{4t_0}t} \right), \label{eq:TL_8b}\\
    E^{c}_{\text kin}(t)/t_0 = -\frac{J}{2t_0}\left(1-e^{-\frac{J^2}{4t_0}t}\right).\label{eq:TL_8c}
    \end{align}
\end{subequations}

In Fig.~\ref{fig:ferro}, we compare the computational results (solid lines) with the TLS results (dashed lines), where one can observe a good agreement for all operators up to $J/t_0 \lesssim 1.5$. Note that as $J/t_0$ increases, oscillations appear. These correspond to magnon reabsorptions and reemissions not captured in the Markovian approximation. Further, the very short-time window is not captured either since there, non-Markovian processes are always present (see, e.g., Refs.~\cite{Buesser2014,deVega2017}).

\subsubsection{Energy transfer and crossover regime $J \sim 2t_0$}

\label{sec:level3.1.1}

\begin{figure}[t]
\includegraphics[width=0.9 \linewidth,, height=6 cm]{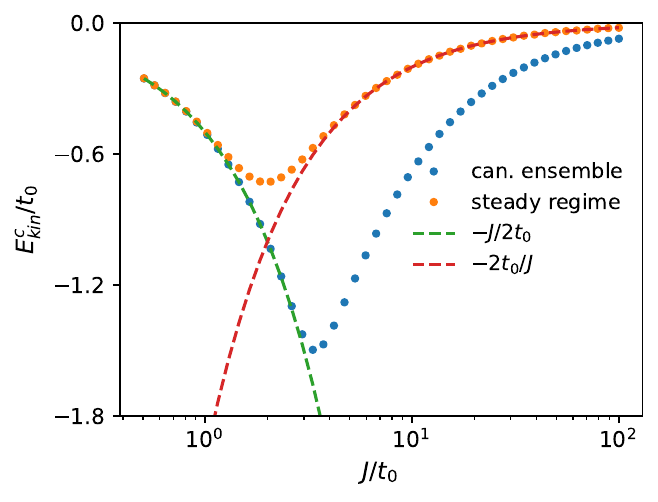}
\caption{ \label{fig:thermo} Steady-state expectation value of the conduction electron kinetic energy as a function of coupling strength $J$, shown on a logarithmic scale. The results are compared with the prediction of the canonical ensemble.
We compute the thermal expectation values on finite systems and
extrapolate to the thermodynamic limit:
A linear regression is used to fit the data in the weak- and 
strong-coupling regimes, while a quadratic regression is applied in the intermediate regime to data from  $L =[108,128,160,180,216,224]$. Dashed lines represent the energy transferred $\Delta E$  from the conduction electron to the localized spin chain for the two different regimes $J/t_0\ll 1$ [green line, $\Delta E =-J/(2t_0)$] and $J/t_0\gg 1$ (red line, $\Delta E = - 2t_0 /J$), respectively.}
\end{figure}

For small $J/t_0$, a complete magnetization transfer  takes place on time scales proportional to $4 t_0/J^2$. At the same time, the kinetic energy of the electron cannot be further reduced after this process since the energy transfer is tied to magnetization transfer. In the small $J/t_0$ regime, the energy transferred $\Delta  E$ to the spin system is proportional to the exchange coupling. However, as shown in Fig.~\ref{fig:thermo}, this linear dependence breaks down when $J/t_0\gtrsim 1$. In this intermediate coupling regime, the momentum transfer approaches $\Delta k \approx \pi/2$ as $J/t_0 \approx 2$.
Thus, the initial state couples to the region of the maximum density of states in the dispersion.
Therefore, as mentioned before, additional effects beyond the simple TLS picture appear, such as repeated absorption and emission of the magnon, the van-Hove singularity at the band bottom becomes increasingly relevant \cite{Henning2012}, and the  overlap of the initial state becomes significant with more than two many-body eigenstates. 
Consequently, the TLS description breaks down and fails to capture the relaxation dynamics for $J\gtrsim t_0$.

For $J>2t_0$, the energy needed for a full spin flip of the conduction electron's spin exceeds the electronic kinetic energy, and thus a complete spin reversal process is not possible.
Figure~\ref{fig:mom_ferro}(c)  shows the quasimomentum occupation $n_{k=\pi/2,\uparrow}$. For $J/t_0=0.9$, this quantity fully relaxes to zero, while for large $J/t_0=3$, there are several oscillations and a non-zero steady-state value.

\subsubsection{\label{sec:level3.1.3} Regime of $J \gg 2t_0$}

The time evolution of expectation values in the strong-coupling limit is illustrated in the right part of Fig.~\ref{fig:ferro}. As the exchange interaction increases, magnon emission and absorption events intensify, resulting in  coherent oscillations seen in all expectation values.

Moreover, the steady values of $S_c^z$ and $S_{cf}$ are far from a complete spin flip and a local spin-spin correlation of one-quarter, respectively. This behavior is corroborated by the spectral decomposition of the initial state  and the energy spectrum discussed in  Appendix~\ref{app:weight}, which suggests an equal contribution from the singlet and the triplet sectors in the initial state for large $J\gg t_0$. 
The expectation of an equal contribution from the singlet and triplet sectors is consistent with the long-time averages of local spin-spin correlations shown in Fig.~\ref{fig:ferro}(e), where increasing $J/t_0$ produces long-time averages around $S_{cf}  \approx -0.25$.

\begin{figure}[t]
\includegraphics[width=0.9 \linewidth, height=6 cm]{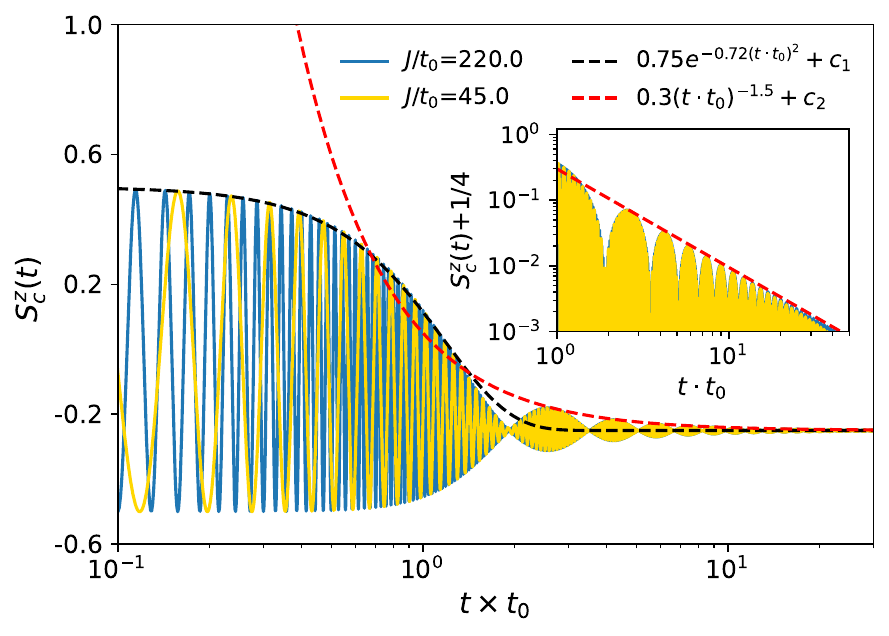}
\caption{\label{fig:large_J}
Time dependence of the conduction electron's spin polarization $S^{z}_c(t)$ for the initial state from Eq.~\eqref{eq:initial_state1} and  large values of the exchange coupling $J/t_0 = 80, 220$. 
Here, we observe a Gaussian  decay for short times and a decay at long times characterized by a power law $\sim  1/t^{3/2}$. The main plot is a  semi-log plot to highlight the exponential decay, while the inset is a log-log plot to visualize the power-law behavior (note that $S_c^z$ is shifted by $0.25$ in the inset). 
The envelopes are fitted to a Gaussian decay at short times
and to a power-law at long times  (see the legend). The values of the fitting parameters of $c_1$ and $c_2$ are $-0.25$. }
\end{figure}

Additionally, the presence of two coherent oscillation frequencies, approximately at $\omega \sim J \pm 2t_0$ (Fig.~\ref{fig:large_J}), indicates that at least three energy levels participate in the dynamics:  the singlet state and two triplet states at $\Delta E = J \pm 2t_0$ relative to the singlet state's energy.

Figure~\ref{fig:large_J} shows the spin polarization of  the conduction electron as a function of time for large $J/t_0$. Due to the presence of two dominant frequencies, the dynamics exhibit a characteristic beating pattern. The envelopes of the time evolution of the spin polarization are the same for both $J/t_0$ values and reveal two distinct temporal regimes, which are plotted with dashed lines. For short times ($t<1/t_0$), the envelope follows a Gaussian profile

\begin{align}
ae^{-b|t\times t_0|^2}
\end{align}
with fitting parameters $a=0.75$ and $b=0.72$. The Gaussian decay indicates a non-Markovian dynamics \cite{deVega2017,coish2006}, which is represented as a black dashed line in Fig.~\ref{fig:large_J}. Such a Gaussian decay is also observed in 
central spin models, where it arises due to a contribution of many individual local environments of localized spins seeing a fluctuating mean-field due the interaction with the central spin \cite{Urbaszek2013,Khaetskii2002,coish2006,Merkulov2002}.

In the long-time regime ($t\gg 1/t_0$), shown in the inset of Fig.~\ref{fig:large_J} using a log-log scale,  the envelope follows a straight line (red dashed line), suggesting a power-law decay proportional to $t^{-3/2}$ due to the van-Hove singularity at the band bottom, in agreement with the exact analytical solution \cite{Henning2012}.

\subsection{Dynamics in the presence of disorder}
\label{sec:disorder}

\begin{figure}[t]
\includegraphics[width=1.0 \linewidth, height=6.0cm]{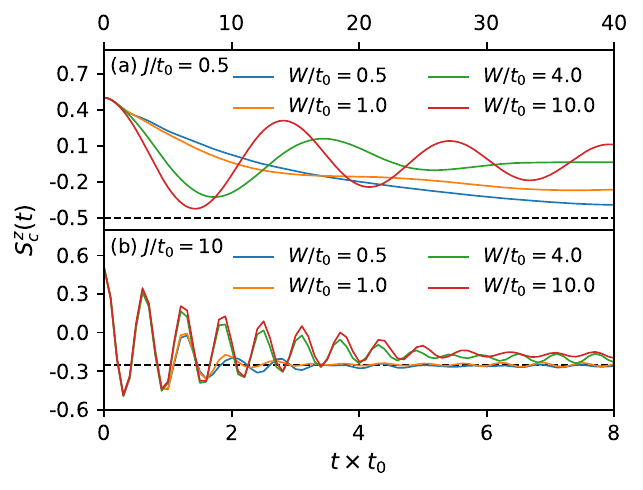}
\caption{\label{fig:disorder}  Time evolution of  the spin polarization $S^z_c$ of the conduction electron  for different disorder strengths ($W/t_0=0.5,1.0,4.0,10.0$) at fixed $J/t_0=0.5$. 
(a) Behavior for a value of the  Kondo-coupling constant $J/t_0=0.5$;  the  dashed line indicates full reversal of the spin orientation to $S_c^z=-0.5$. (b) Results for a larger coupling constant $J/t_0=10.0$; the  dashed line indicates
$S_c^z=-0.25$.
The spin polarization is scaled by the system size, and time is measured in units of $1/t_0$. We use the Krylov method with $L=100$, $k_0=\pi/2$, and an initial ferromagnetic configuration for the localized electrons. 
We use a minimum of 1216 iterations for large $J/t_0$ and a minimum of 2560 iterations for small $J/t_0$.}
\end{figure}

We next demonstrate that adding diagonal disorder suppresses the relaxation
dynamics significantly \cite{nandkishore15,sierant25}. We first consider $J/t_0=0.5$ and then compute the decay of the 
spin polarization for increasing values of disorder strength $W$. Results obtained
from averaging over at least 1216 disorder realizations are shown in Fig.~\ref{fig:disorder}(a)
for $L=80$.

\begin{figure}[t]
\includegraphics[width=1.0 \linewidth, height=5.5cm]
{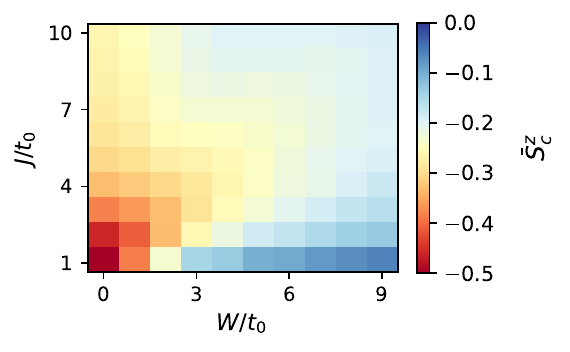}
\caption{\label{fig:disorder2} Time-average $\overline{S}^z_c$ of $S^z_c(t)$ after the fast initial relaxation as a function of the Kondo coupling ($J/t_0$) and disorder strength ($W/t_0$) for $L=80$. Red color indicates a spin-down orientation, while blue corresponds to a vanishing spin polarization. The time average is computed from the time window $[25/t_0,35/t_0]$ with a time step of $\Delta t=0.2/t_0$.}
\end{figure}

While at small disorder, the curves still approach $S_c^z=-1/2$, already at $W/t_0=4$,
the system settles into a state with only a partial magnetization transfer, accompanied by
large oscillations. For a large value of $J/t_0=10$, the spin polarization decays to $S_c^z\approx -1/4$,
yet with large oscillations.
This behavior is summarized in Fig.~\ref{fig:disorder2}, where we plot the  average long-time value of $S_c^z(t)$ as a function of $J/t_0$ and $W/t_0$. Overall, disorder suppresses the relaxation and prevents the magnetization transfer. Note that at small
disorder $W/t_0$, finite-size effects 
are significant, yet as $L$ increases, a complete relaxation is observed for, e.g., $W/t_0=0.5$ (results not shown here). 

Generically, we expect Anderson localization for a charge carrier in one dimension, yet coupling to a many-body bath may lift this \cite{Bonca2018,Krause2021,Sierant2023}. 
Our data are consistent with localization on finite systems for sufficiently large $W/t_0$, as disorder detunes
possible transitions via the coupling to the spin bath. In Fig.~\ref{fig:disorder}(b), we show the spin polarization in the large $J/t_0$ limit, where a simple picture emerges. The electron localizes in a site $i_0$ and the magnetization is coherently transferred back between the localized spin
and the electronic spin.
This picture is supported by the following argument:
The oscillations at large $J/t_0$ are controlled by a frequency $\omega=J$, while the beating behavior 
observed without disorder is increasingly suppressed.

\subsection{Absence of thermalization}
\label{sec:thermalization}

\subsubsection{Absence of thermalization in the $S^z_{\text{total}}=(L-1)/2$ sector}
\label{sec:thermo}

We return to the case of clean systems with $W=0$.
In the subspace with $S^z_{\text{total}}=(L-1)/2$, the Hilbert-space scales only polynomially with system size $L$. Hence, on general grounds, the density of states will not become dense enough to allow for thermalization. Consistently, we find that 
(i) the quasimomentum distribution maintains two sharp finite-momentum peaks for long times and for small and intermediate values of $J/t_0$ [see Fig.~\ref{fig:mom_ferro}(a) and \ref{fig:mom_ferro}(b)], and (ii) the long-time expectation values do not agree with the 
thermal expectation values obtained from the canonical ensemble [see Fig.~\ref{fig:thermo}], which is the most evident for  intermediate values of 
$J/t_0$.

This behavior of $n_k$ is notably different from the one observed for the
Holstein polaron in one dimension, where a thermal-like stationary $n_k$ with a broad momentum distribution centered around $k=0$ is quickly established for a comparable initial condition \cite{Dorfner2015}. The Holstein-polaron chain, remarkably, exhibits clear eigenstate thermalization behavior already on small system sizes \cite{Jansen2019,Schoenle2021}. However, for the Holstein polaron, the Hilbert-space  scales exponentially with system size for the cases studied in Refs.~\cite{Dorfner2015,Jansen2019,Schoenle2021}.

For our spin-polaron model and in the subspace with $S^z_{\text{total}}=(L-1)/2$, this is not the  case. The simplest measure for quantum chaos is the gap ratio \cite{Oganesyan2007}, defined as follows. Let $E_n$ be the many-body eigenenergies ordered as $E_{n+1}> E_n > E_{n-1} > \dots $, then the nearest-neighbor level spacing is defined as 
\begin{align}
\delta_n = E_{n+1}-E_n \,.
\end{align}
The gap ratio is
\begin{align}
r_n = \frac{\text{min}(\delta_n,\delta_{n-1})}{\text{max}(\delta_n,\delta_{n-1})}\,.
\end{align}
Note that we include 50\% of all eigenstates centered around the mean energy in the averaging procedure.
We find that the average gap ratio $r=\langle r_n\rangle$ in the $S^z_{\text{total}}=(L-1)/2$ is far away from GOE behavior
expected for a quantum chaotic many-body system \cite{DAlessio2016}
(data not shown in the figures).

\section{Decay of several charge carriers in a ferromagnetic background}
\label{sec:many-particles}
\subsection{Two-electron case}

The first step consists in going to the two-electron case moving in a ferromagnetic background, studied extensively in the literature.
We aim at using the quasimomentum distribution function to illustrate the expected absence of thermalization in this few-body problem.
Two initial states are considered: (i) two electrons with spin up and (ii) two electrons with opposite spin.
In both cases, one electron is initialized with  
quasimomentum $|k_0=\pi/2\rangle$ and the other one at $|k_0=-\pi/2\rangle$
The corresponding initial states are:

\begin{align}
|\Psi_{0}^{(\uparrow\uparrow)} \rangle &= 
\hat c^{\dagger}_{-\pi/2,\uparrow} \hat c^{\dagger}_{\pi/2,\uparrow}
| 0 \rangle_c  \otimes |\text{FM} \rangle_f, 
\label{eq:2N_upup} \\[6pt]
|\Psi_{0}^{(\uparrow\downarrow)} \rangle &= 
\hat c^{\dagger}_{-\pi/2,\downarrow} \hat c^{\dagger}_{\pi/2,\uparrow}
| 0 \rangle _c \otimes
|\text{FM}\rangle_f.
\label{eq:2N_updown}
\end{align}

\begin{figure}[t]
\includegraphics[width=1.05 \linewidth, height=7.20 cm]{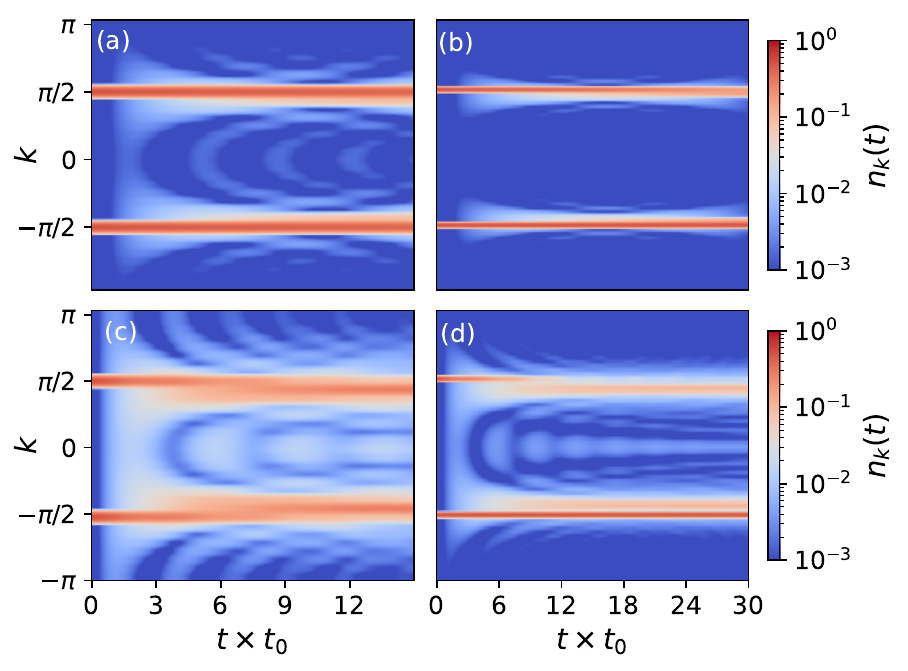}%
\caption{\label{fig:two-ele} Relaxation dynamics of the two-electron Kondo model. In (a) and (c), we present the time evolution of the quasimomentum distribution $n_k(t)$ for the parallel-electron initial state with two electrons with spin up [see Eq.~\eqref{eq:2N_upup}] with system size $L=32$, and $J/t_0=0.4$ and $1.0$, respectively. As time evolves, the quasimomentum distribution develops two narrow peaks located around $k =\pm \pi/2 \mp J/(4t_0)$. (b) and (d) show the  corresponding time evolution for the  initial state with electrons of opposite spin orientation [see Eq.~\eqref{eq:2N_updown}] with $L=64$, and $J/t_0=0.4$ and $1.0$, respectively. In this case,  a faster decay of the spin-up electron compared to the spin-down electron is evident. }
\label{fig:momentum_two_particle} 
\end{figure}

Our results for the quasimomentum distribution in the two-electron case are shown in Fig.~\ref{fig:two-ele}. In the first case, both electrons independently emit one magnon,
exhibiting a dynamics that is overall similar to the one-electron case. 
Essentially, $n_k(t)$ exhibits   two well-defined peaks at $k=-\pi/2 + J/(4t_0)$, and $k=\pi/2 - J/(4t_0)$.  As time evolves, the function decays as an exponential function with a constant 
that is slightly renormalized from $J^2/4$.
Note that in this 
four-body problem of two electrons and two magnons, 
interactions  lead to a mid-spectrum bound state, studied in Refs.~\cite{Rausch2019,Moeller2012}.

In the second case, the up electron emits a magnon that can be reabsorbed by the down electron, leading to more interesting dynamics in the effective three-body problem.
The spin-down electron momentum occupation at $k_0=\pi/2$ does essentially not decay, while the one of the 
spin-up electron decays. Notably, a small
weight emerges at zero momentum,
reminiscent of a bound state \cite{Rausch2019,Moeller2012,Moeller2012a}.
The exponential decay seen in the 
one-electron case gets more modified
due the more complicated structure of decay channels in the two-particle case \cite{Moeller2012,Rausch2019}.
In both cases, there is clearly
no relaxation to a thermal state, as expected.

\subsection{Towards a finite density of charge carriers}
We now go to a finite filling of one quarter. Here, the accessible system sizes are quickly rather restricted.
Specifically, we consider  the initial state
\begin{align}
|\Psi_0 \rangle = \hat c^{\dagger}_{-3\pi/4,\uparrow} \hat c^{\dagger}_{3\pi/4,\uparrow}
\hat c^{\dagger}_{\pi/2,\uparrow}
\hat c^{\dagger}_{\pi/4,\uparrow} 
|0 \rangle_c \otimes | \text{FM} \rangle_f \label{eq:Ne4} \,.
\end{align}

\begin{figure}[t]
\includegraphics[width=1.05 \linewidth, height=8.2 cm]{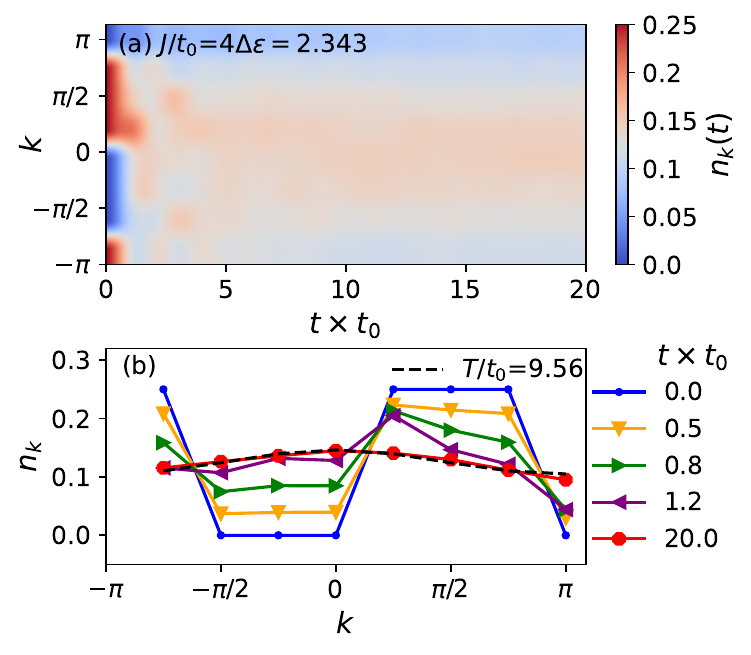}%
\caption{Relaxation dynamics in the  Kondo model at quarter filling. (a) Time evolution of the quasimomentum distribution $n_k(t)$ for the initial state from Eq.~\eqref{eq:Ne4} with system size $L=8$ and $J/t_0=2.343$. During the relaxation process, a maximum in $n_k(t)$  forms at $k=0$, indicating the stationary regime. Panel (b) presents several  cuts through panel (a) at fixed times, specifically showing $n_k(t)$ at $t \, t_0=0.0,0.5,0.8,1.2$ and $20.0$. Additionally, the black dashed line corresponds to the expectation value of the quasimomentum distribution in the canonical ensemble at a $T=9.56\,t_0$.  $\Delta \epsilon$ is the minimum energy of one particle to jump from one state $k_n$ to $k_{n+1}$, where $\epsilon_{k_{n+1}}<\epsilon_{k_{n}}$. Thus $\Delta \epsilon= \epsilon_{2\pi/L} - \epsilon_{k=0}$.}

\label{fig:momentum_L8_N4} 
\end{figure}

We display numerical results for 
$N_e=4$ in $L=8$ sites in Fig.~\ref{fig:momentum_L8_N4}.
 Even for this modest 
case of $L=8$ sites, a symmetric quasimomentum distribution centered around $k=0$ emerges relatively quickly.
Figure~\ref{fig:momentum_L8_N4}(b) 
shows $n_k(t)$ versus $k$ for a number of different times, where
at $t=20/t_0$, the distribution has become practically symmetric.
Moreover, the long-time form of 
$n_k(t)$ agrees very well with the thermal expectation values computed for the chosen initial condition 
[dashed line in Fig.~\ref{fig:momentum_L8_N4}(b)].

\begin{figure}[t]
\includegraphics[width=1.0 \linewidth, height=6.0cm]{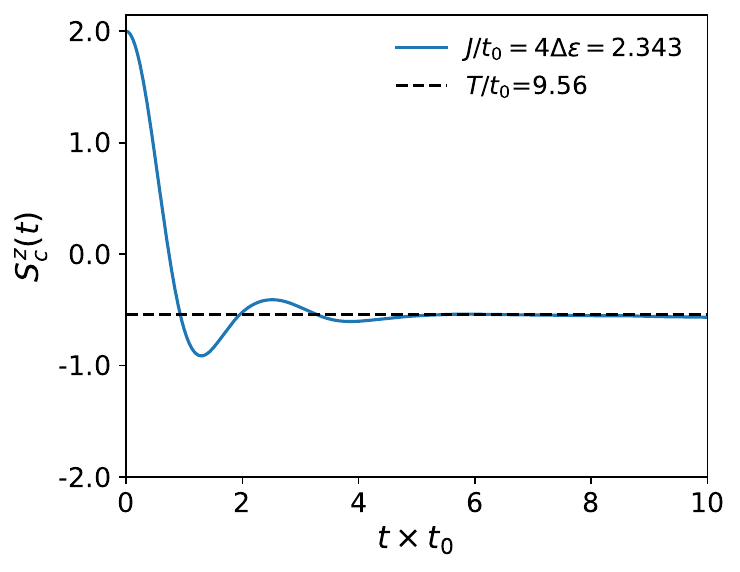}%
\caption{Time evolution of the conduction-electron spin polarization $S^{z}_{c}$ (solid line) for the initial state from 
Eq.~\eqref{eq:Ne4} for system size $L=8$ and density $n=1/4$. The steady regime occurs at $ t\times t_0 \approx 6$, where  $S^z_c$ exhibits small oscillations around the expectation value of spin polarization in the canonical ensemble at a $T=9.56t_0$.
}
\label{fig:S_z_L8_N4} 
\end{figure}

The same holds true for the decay of the total spin polarization of the conduction electrons, plotted in Fig.~\ref{fig:S_z_L8_N4}. $S_c^z(t)$
quickly decays and approaches a stationary value after a short time.
The decay shows deviations from the
exponential decay seen for one charge carrier due to the additional bands
and the bound state in the spectrum \cite{Rausch2019,Moeller2012}.
This stationary value is surprisingly close to the thermal expectation value (dashed line).
To summarize, in this section, we numerically demonstrated that a system with just four electrons and four emitted magnetic excitations
exhibits dynamics consistent with 
thermalization.

The results of this section suggest that at any finite electronic density, the dynamics will be consistent with thermalization. While hard to access numerically, we expect that this holds even in the low-density limit above the ferromagnetic ground state of the model. Understanding the dependence of thermalization time scales on electronic density is an interesting question that warrants a separate study with different numerical tools such as matrix-product state methods \cite{schollwock2011}.

\section{Decay of a single charge carrier in a correlated and antiferromagnetic background}

\label{sec:AFM}

We now demonstrate that for the Kondo-Heisenberg model and for subspaces whose dimension scales exponentially with system size, a relaxation to the thermal state can be spotted in finite-size data.

\begin{figure}[t]
\includegraphics[width=0.95 \linewidth, height=5.5 cm]{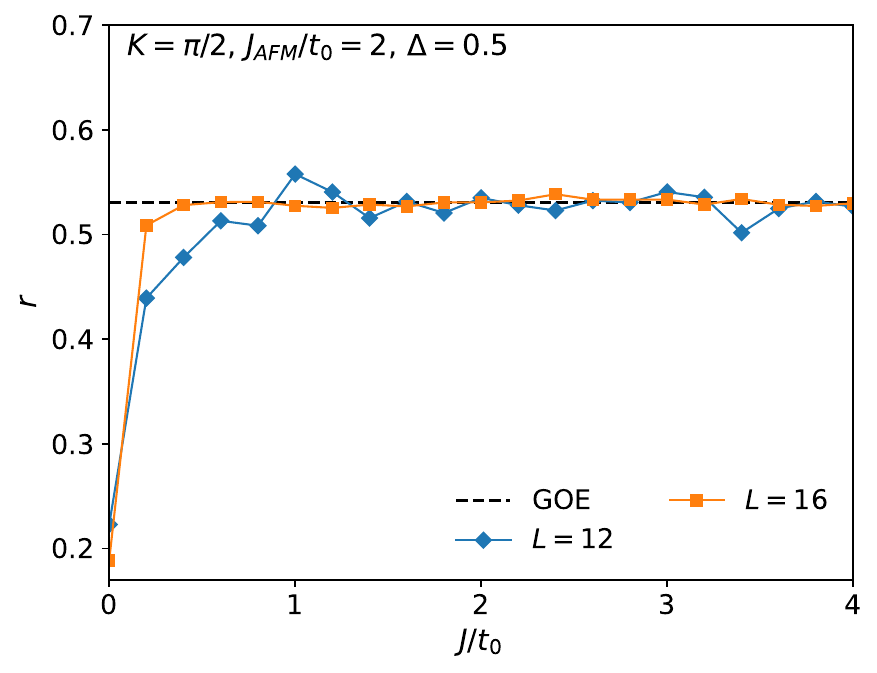}%
\caption{Average gap ratio $r$ of the model defined in Eq.~\eqref{eq:kondo} with $J_{\text{AFM}}/t_0=2$, evaluated in the $K=\pi/2$, $N_e=1$, and $S^z_{\text{total}}=1/2$ sector, as a function of the Kondo coupling strength for two different system sizes $L=12,16$. The spin exchanges of all spin-spin interactions contain an anisotropic spin exchange $\Delta=0.5$ [see Eq.~\eqref{eq:delta]}]. Here, we restrict our attention to the many-body eigenstates that are located in the middle half of the energy-ordered list of states for their sample, thereby avoiding high- and low-energy states. The dashed line represents the value of $r_{\text{GOE}}$ taken from \cite{DAlessio2016}. 
}
\label{fig:gapratio} 
\end{figure}

To that end, we turn to $J_{\text{AFM}}>0$ and we first analyze the gap ratio as
a function of $J/t_0$ and  $J_{\text{AFM}}/t_0$, shown in Fig.~\ref{fig:gapratio}.
There, we introduce an anisotropy in all
spin exchange terms in the Hamiltonian:
\begin{align}
\hat{\boldsymbol{S}}_i \cdot \hat{\boldsymbol{S}}_j
 \rightarrow \frac{1}{2} (S_i^+ S_j^- + \mathrm{H.c.}) + \Delta S^z_i S^z_j\,, \label{eq:delta]}
 \end{align}
which breaks SU(2) symmetry. The data 
shows that the finite-size data converge quickly to the expected GOE value. This holds true for arbitrary values of $J_{\text{AFM}}\geq 0$, and thus a dispersion or an interaction in the system of localized spins is not essential to obtain GOE behavior.

Based on the analysis of the gap ratio, we choose as model parameters $J_{\text{AFM}}/t_0 =2$, $J/t_0=2$, and $\Delta = 0.5$. As the initial state, we select one of the  quasimomentum states in the $K=\pi/2$ sector. Specifically, the state reads
\begin{align}
|\Psi_{0,\text{KHM}} \rangle =\sum^{L}_{j=1} \frac{e^{-i\frac{\pi}{2} j}}{\sqrt{L}} \hat{T}^{j}\left(\left|\uparrow\right>_c  \otimes  \left|\uparrow \downarrow \uparrow \downarrow \dots\right>_f\right),
\label{eq:initial_state_AFM}
\end{align}
where the spin pattern continues periodically until the full system size, $\hat{T}$ is the generalized translation operator. 
The excitation energy $\langle \Psi_{0,\text{KHM}}| \hat H| \Psi_{0,\text{KHM}}\rangle - E_0$ corresponds to a finite temperature
in a canonical ensemble.

\begin{figure}[t]
\includegraphics[width=1.0 \linewidth, height=7 cm]{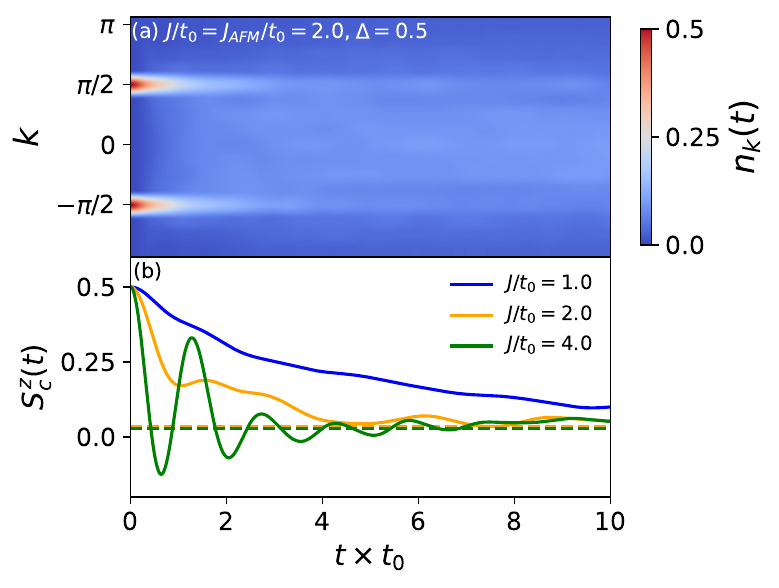}%
\caption{ Relaxation dynamics in the Kondo-Heisenberg model. In (a), we show $n_k(t)$ as a function of quasimomentum $k$ and time  for the initial state from Eq.~\eqref{eq:initial_state_AFM} with $L=16$, $J_{\text{AFM}}/t_0 = 2.0$, $J/t_0 = 2.0$, and $\Delta = 0.5$. As time progresses, the quasimomentum distribution develops a broad peak centered around $k = 0$.  (b) Time evolution of the conduction electron's polarization $S^{z}_{c}$ (solid line) for the same initial state,  compared to its thermal expectation value from the canonical ensemble (dashed lines) for $J_{\text{AFM}}/t_0=2.0$ and $\Delta = 0.5$, and different values of $J/t_0$. Specifically, for $J/t_0 = 1.0,\, 2.0,$ and $4.0$, we evaluated the canonical
ensemble expectation values at the temperatures consistent with the initial state, which yields
$T = 3.096\,t_0$, $3.112\,t_0$, and $3.375\,t_0$, respectively.
We observe a reasonably  good agreement at long times, given the small system sizes.
}
\label{fig:therma_L16} 
\end{figure}

Figure~\ref{fig:therma_L16}(a) shows the quasimomentum distribution for the relaxation dynamics starting from Eq.~\eqref{eq:initial_state_AFM} for the selected parameters.
In the initial state, the electron has weight at $k_0=\pi/2$ and $k_0=-\pi/2$. These peaks fade away, while a single peak at zero momentum appears and  becomes the single maximum at the longest simulated times. 

In Fig.~\ref{fig:therma_L16}(b), we vary $J/t_0$ and plot the time evolution of $S_c^z$. While the relaxation dynamics clearly depend on $J/t_0$, for all selected parameters, the curves approach the thermal expectation value expected for the initial conditions. Obviously, due to the small system size, there are remaining oscillations, yet the overall trend is compatible with the expectations from the analysis of the gap ratio, that is, thermalization from the closed quantum system perspective \cite{DAlessio2016}.

\section{\label{sec:level4}Summary and conclusion}

In this work, we conducted a comprehensive analysis of the non-equilibrium dynamics and thermalization of charge carriers in the  Kondo lattice chain. 
We focus on initial states that are product states between the conduction electron sector and the one of the localized spins.
The study is carried out using the time-dependent Lanczos method and exact diagonalization.

We first provided a discussion of the well-studied case of one or two charge carriers in 
in a fully polarized spin background from the thermalization perspective. While exact results exist (see, e.g., Ref.~\cite{Henning2012}), a generic picture arises from 
considering the relevant electronic levels 
(in the case of one charge carrier, just two)
coupled to a bosonic bath. Utilizing standard approaches, this generically leads to an exponential decay, in agreement with exact results \cite{Henning2012} and with our exact numerical results. Furthermore, we investigated the effect of disorder on the 
relaxation dynamics, where the behavior at large values of disorder strength is consistent
with localization.
Most notably in the context of the current study, the expected absence of thermalization is evident in the stationary form of the quasimomentum distribution for one or two charge carriers that exhibit resilient finite-momentum peaks.

As our main results, we considered two modifications of the initial state: either by going to a finite density of charge carriers in a ferromagnetic background or by injecting 
one electron into an antiferromagnetic background, the time evolution of observables becomes consistent with thermalization.
This is corroborated by a direct comparison with expectation values computed in a compatible canonical ensemble and an analysis of the gap ratio.

An interesting open question concerns the dependence of thermalization and relaxation time scales on electronic density in the ferromagnetic regime or in the vicinity of half filling at low energy densities. Generally, we expect dynamics compatible with thermalization as long as the model is generic (i.e., not integrable) and as long as the dynamics occurs in a macroscopically large subspace. This question could be addressed with matrix-product-state methods (see, e.g, Ref.~\cite{schollwock2011}).

As an outlook, we envision to add optical phonons to the problem, which would allow us to study the competition between two bosonic baths. Technically, this could be simulated with advanced matrix-product state
methods for phonons \cite{Brockt2015,Jansen2020,Jansen2023,Koehler2021} or a classical treatment of the phonons in the spirit of multitrajectory Ehrenfest dynamics \cite{tenBrink2022} or the truncated Wigner approximation (see, e.g., Ref.~\cite{Paprotzki2024}). Moreover,
the initial state can be prepared as the result of an explicit modeling 
of an optical pulse.\\

\section*{Data Availability}

The data that support the findings of this article are openly available \cite{P-R_2025}. Data that are not openly available, are available from the authors upon reasonable request.

\acknowledgements
We thank A. Honecker, E. Jeckelmann, S. Kehrein, and S. Manmana for useful discussions.
This work was funded by the Deutsche Forschungsgemeinschaft (DFG, German Research Foundation) --  217133147,
436382789, 493420525; via CRC 1073 and large-equipment grants (GOEGrid).

\appendix

\section{Spectral decomposition of the initial state with one charge carrier in the fully polarized background}
\label{app:weight}

Even within the same momentum sector (in our case, $K=\pi/2$), the initial state from Eq.~\eqref{eq:initial_state1} remains significantly distinct from the ground state. One way to illustrate this separation between the initial state and the ground state is to expand the initial state (for the sector $K=\pi/2$) into contributions from the   ground state and excited states. We start defining the spectral weight in the ground state $\left|\Psi_{\text{gs}}\right>$ as $P_{\text{gs}}=|\left<\Psi_{0,F}|\Psi_{\text{gs}}\right>|^2$, and the total spectral weight in the excited states as $P_{\text{ex}}=\sum_{n>0}|\left<\Psi_{0,F}|\Psi_n\right>|^2$. These weights satisfy the normalization condition: $P_{\text{gs}} +P_{\text{ex}} = 1$. To further characterize the initial state, we also evaluate  the spectral-weighted average of the local spin-spin operator, which reduces to:

\begin{equation}  
\begin{split}
K_{cf}(\Psi_{\alpha}) & =\sum_n \left<\Psi_{n}|\Psi_{\alpha}\right>^2
\left<\Psi_{\alpha}\right|
{\hat{\textbf{S}}}_f\cdot 
{\hat{\textbf{S}}}_c \left|\Psi_{n}\right>\\
\Rightarrow K_{cf}(\Psi_{\text{gs}}) &=\left<\Psi_{\text{gs}}\right|{\hat{\textbf{S}}}_f\cdot{\hat{\textbf{S}}}_c \left|\Psi_{\text{gs}}\right>.
\end{split} \label{eq:crazyK}
\end{equation}

\begin{figure}[t]
\includegraphics[width=1.0\linewidth, height=5.5cm]{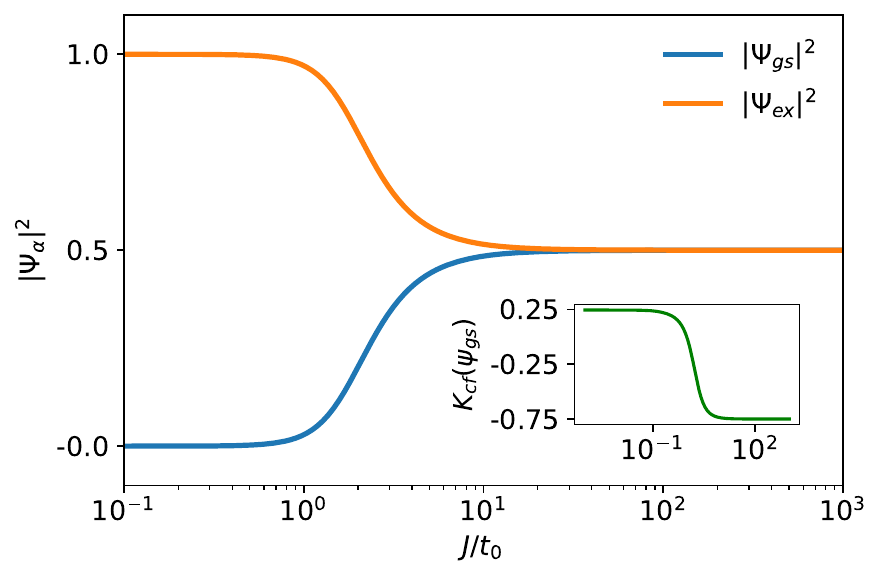}%
\caption{\label{fig:Weight} Spectral decomposition $|\Psi_{\alpha}|^2$ of the initial state from Eq.~\eqref{eq:initial_state1} into ground ($\alpha=\text{gs}$) and excited ($\alpha=\text{ex}$) states for the $K=\pi/2$ and $S^z_{\text{total}}=(L-1)/2$ subspace with $L=126$. In the inset, we plot the expectation value of $K_{cf}$ in the ground state for $K=\pi/2$, see  Eq.~\eqref{eq:crazyK}. 
}
\end{figure}

Figure~\ref{fig:Weight} shows the spectral weight distribution of the  initial state. For a strong Kondo coupling, the initial state has an equal contribution from the ground state and excited states. For arbitrary values of $J/t_0$, since the local spin-spin correlation in the ground state stems from the singlet contribution (as shown in the inset of Fig.~\ref{fig:Weight}), the contribution from the singlet decreases away from the limit $J/t_0\to \infty$. 
By decreasing $J/t_0$, the spectral weights rearrange such that the spectral weight increases in the manifold of excited states. 

Calculating the spectral-weighted average $K_{cf}$ from Eq.~\eqref{eq:crazyK} for the ground state and excited states reveals that, for large $J/t_0$, these approach:
\begin{align}
K_{cf}(\Psi_{\text{gs}})= -\frac{3}{4} +\frac{t_0^2}{J^2} \,
\end{align}
and
\begin{align}
K_{cf}(\Psi_{\text{ex}})= \frac{1}{4} +\frac{1}{L}\frac{t_0^2}{J^2} \,,
\end{align}
respectively. Moreover, the following dependence of $S_{cf}$ on $J/t_0$  emerges in this limit in the steady state:
\begin{align} 
    S_{cf}(t \rightarrow \infty) &=P_{gs}K_{cf}(\Psi_{\text{gs}}) +P_{ex}K_{cf}(\Psi_{\text{ex}})\nonumber \\
    &=-\frac{1}{4}+\left(2+
    \frac{1}{2L}\right)\frac{t_0^2}{J^2}+\mathcal{O}(J^{-4})
    \,.
\end{align}

\begin{figure}[t]
\includegraphics[width=1.0 \linewidth, height=6cm]{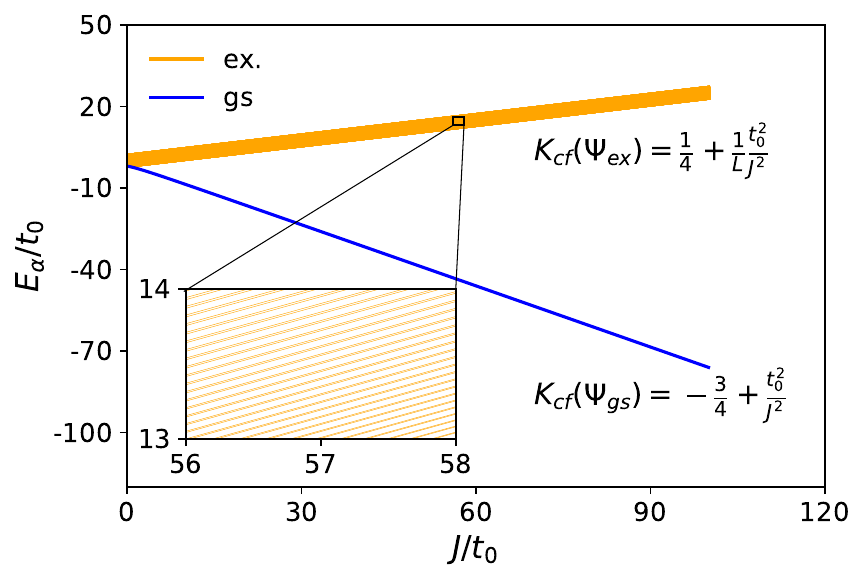}
\caption{\label{fig:EvsJ}  Energy spectrum as a function of the Kondo exchange coupling $J$ for a system  size $L=216$ in the $K=\pi/2$ sector. 
The thin blue line is the ground-state energy, the thick orange line represents the band of excited states (see the inset for a zoom into that region).
For any finite $J/t_0$, the band of excited states is separated from  the ground state by an excitation gap of $\mathcal{O}(J)$. By calculating the spectral-weighted average of the local spin-spin
operator from Eq.~\eqref{eq:crazyK}, we observe that the ground state is a singlet  [$K_{c,f}(\Psi_{\text{gs}})$] and the excited states are in the triplet sector [$K_{c,f}(\Psi_{\text{ex}})$] for large $J/t_0$, as expected \cite{ueda1991}.}
\end{figure}

Analogously, we find for the kinetic energy:
\begin{align}
E^c_{\text{kin}}(t \rightarrow \infty)=-\frac{2t_0^2}{J} + \frac{1}{L}\frac{t_0^2}{J}\,.
\end{align}
In  Fig.~\ref{fig:EvsJ}, we observe a gap between ground state and excited states, which is proportional to the exchange coupling \cite{ueda1991}. 

\section{Two-level plus bath approximation}
\label{app:markov}

We briefly sketch the main approximations used in the analytical
approach that leads to the exponential decay discussed in Sec.~\ref{sec:level3.1.1}, following Ref.~\cite{seke1983} and \cite{berman2010}.
The interaction between the two-level system and the bath is 
characterized by the coupling constant $\gamma_{q}$, with $q$ representing the  wavevector of the emitted excitation in one dimension. In general, the tensor can be defined as:
\begin{align}
\gamma_q= -i\omega f(\omega_q)\left<e\right|\hat{\textbf{O}}_{TLS}\left|g\right>\cdot\mathbf{e}_q \,,
\end{align}
where $\omega$ is the energy difference between the excited and ground states, $f(\omega_q)$ is the  density of states at the energy of the  emitted excitation, $\hat{\textbf{O}}_{TLS}$ is the operator that couples the TLS to the bath, and $\mathbf{e}_q$ is the polarization vector associated with the emitted excitation, orthogonal to $q$.
To study the relaxation dynamics, we focus on the time evolution of the population inversion operator 
\begin{align}
\sigma_{\textsl{e}\rightarrow \textsl{g}}=(\left|\textsl{g}\right> \left<\textsl{g}\right|-\left|e\right> \left<e\right|) \,,
\end{align}
 which tracks the transition probability between the two levels. 
 A standard method to derive its dynamics is to use the Robertson projection operator technique \cite{robertson1966,seke1980}, which projects the full density matrix evolution onto the relevant subspace spanned by $\sigma_{e\rightarrow \textsl{g}}$. This results in an integro-differential equation involving memory kernels and the expectation value of $\sigma_{e\rightarrow \textsl{g}}$ at earlier times. However, solving this equation exactly is often intractable. To proceed, we assume Markovian dynamics (justified when the bath correlations decay rapidly), allowing us to neglect memory effects and replace the history-dependent term by its instantaneous value. This approximation simplifies the equation to a first-order differential equation. Finally, by taking the thermodynamic limit (replacing sums over $q$ by an integral), 
 and introducing an exponential cutoff in time (see   Ref.~\cite{seke1983} for more detail), we obtain an exponential decay of the excited-state population, a hallmark of Markovian relaxation in open quantum systems.

In the context of this study, we compute the transition probability between states $|e\rangle = |k_0,\uparrow \rangle$ and $|\textsl{g}\rangle=|\pm (k_0 -\Delta k),\downarrow\rangle$, representing the excited state $\left|e\right>$ and the degenerate ground state $\left|\textsl{g}\right>$, respectively. The electronic occupation of a
spin-up electron is given by: 
\begin{equation} \label{eq:TL_1}
\left| n_{\uparrow} (t) \right> = b_2(t)e^{-i\omega_0 t}\left|2,0\right> + \int_{-\infty}^{\infty}d\omega_q \,b_{1,q}(t)e^{-i\omega_q t}\left|1,\omega_q\right>\,.
\end{equation}
Here, the ket $\left|2,0\right>$ denotes the state of the conduction electron  with kinetic energy $\varepsilon_{2}=-2t_0\cos(\pi/2)=0$ and no magnons in the localized spin chain, while $\left|1,\omega_q\right>$ represents the state of the conduction electron with kinetic energy $\varepsilon_{1}=-2t_0\cos(\pi/2-\Delta k)\approx-J/2$ (two-fold degenerate) and a magnons contained in the  spin chain, with energy $\omega_q/2$. 
Here, $b_2(t)$ and $b_{1q}(t)$ represent the probability amplitudes for the kets $\left|2,0\right>$ and  $\left|1,\omega_q\right>$, respectively, and $\omega_0$ is the energy transferred to the bath. Due to the initial condition, the system is prepared  completely in the excited state $|2,0\rangle$ at $t=0$,  implying that:
\begin{equation} \label{eq:TL_2}
b_2(t=0) = 1, b_{1q}(t=0) = 0\,.
\end{equation}
To characterize the interaction between the TLS and the bosonic bath, we need to solve a set of coupled differential equations given by:
\begin{subequations}
    \begin{align}
        \dot{b}_2 &= -i\sqrt{\frac{\gamma }{\pi}}\int_{-\infty}^{\infty} d\omega_q \, f(\omega_q) e^{-i(\omega_q-\omega_0)t}b_{1q},  \label{eq:TL_3a}\\
        \dot{b}_{1q} &= -i\sqrt{\frac{\gamma }{\pi}} f(\omega_q) e^{-i(\omega_q-\omega_0)t}b_{2}\,, \label{eq:TL_3b}
    \end{align}
\end{subequations}

where the parameter $2{\gamma}$  characterizes the decay rate of the excited state in the Markovian limit. The function $f(\omega_q)$ is a real and dimensionless quantity that expresses the frequency dependence of the magnon density states. To arrive 
at Eq.~\eqref{eq:TL_3a} and Eq.~\eqref{eq:TL_3b}, we assumed the rotating-wave approximation (RWA) \cite{cohen1992}, which neglects rapidly oscillating counter-rotating terms. Now, at short times 
($|\omega_q-\omega_0| t\ll 1$), if we consider $|\omega_q-\omega_0|>\gamma$, we can expand the exponential factor in Eq.~\eqref{eq:TL_3a} in a Taylor series in $\gamma$. This yields an approximate expression for the amplitudes in order $\mathcal{O}(\gamma)$:
\begin{equation} \label{eq:TL_4}
b_2(t) \sim b_2(0)\left(1- \frac{\gamma F}{2 \pi}t^2\right)+\dot{b}_2(0)\left(t- \frac{i\bar{\omega}}{2}t^2\right)\,,
\end{equation}
where $F$ is the spectral weight, given by:
\begin{equation} \label{eq:TL_5}
F = \int_{-\infty}^{\infty}[f(\omega_q)]^2 d\omega_q
\end{equation}
and 
\begin{equation} \label{eq:TL_6}
\bar{\omega} = \frac{1}{F}\int_{-\infty}^{\infty}d\omega_q(\omega_q-\omega_0)[f(\omega_q)]^2 \,.
\end{equation}
Inserting the initial state from  Eq.~\eqref{eq:TL_2}, we can simplify the expression Eq.~\eqref{eq:TL_4} to
\begin{equation} \label{eq:TL_7}
b_2(t) \sim 1-\frac{\gamma F}{2\pi}t^2.
\end{equation}

\bibliography{apssamp_V3}

\end{document}